\documentclass[prb,reprint,a4paper,longbibliography,superscriptaddress,english,showpacs]{revtex4-1} 

\usepackage{xcolor}
\usepackage{graphicx} 
\usepackage{amsmath}
\usepackage{amssymb}
\usepackage[normalem]{ulem}
\usepackage{todonotes}

\definecolor{darkblue}{rgb}{0.1,0.2,0.6} \definecolor{darkred}{rgb}{0.8,0.1,0.2}
\usepackage[colorlinks,citecolor=darkblue,linkcolor=darkred,urlcolor=darkblue]{hyperref}
\usepackage[all]{hypcap} 

\newcommand{\bra}[1]{\langle #1|} 
\newcommand{\ket}[1]{|#1\rangle}

\newcommand{\E}{\mathrm{e}} 
\newcommand{\I}{\mathrm{i}}

\newcommand{\ie}{\textit{i.e.} } 
\newcommand{\eg}{\textit{e.g.} }

\newcommand{\up}{\uparrow} 
\newcommand{\dw}{\downarrow} 
\newcommand{\tr}{\mathrm{Tr}}

\graphicspath{{./}{./images/}}

\begin{document}
\title{Operator entanglement entropy of the time evolution operator in chaotic systems}

\author{Tianci Zhou}
\affiliation{Institute for Condensed Matter Theory and Department of Physics, University of Illinois at Urbana-Champaign, Urbana, IL 61801, USA}
\email{tzhou13@illinois.edu}
\author{David J. Luitz}
\affiliation{Institute for Condensed Matter Theory and Department of Physics, University of Illinois at Urbana-Champaign, Urbana, IL 61801, USA}
\email{dluitz@illinois.edu}
\date{\today}

\begin{abstract} 
  We study the growth of the operator entanglement entropy (EE) of the time evolution operator in
  chaotic, many-body localized (MBL) and Floquet systems. In the random field Heisenberg model we
  find a universal power law growth of the operator EE at weak disorder, a logarithmic growth 
  at strong disorder, and extensive saturation values in both cases. In a Floquet spin model,
  the saturation value after an initial linear growth is identical to the value of a random unitary
  operator (the Page value). We understand these properties by mapping  the operator EE to a global
  quench problem evolved with a similar parent-Hamiltonian in an enlarged Hilbert space with the
  same chaotic, MBL and Floquet properties as the original Hamiltonian. 
  The scaling and saturation properties reflect the spreading of the state 
  EE of the corresponding time evolution. We conclude that the EE of the evolution operator should 
  characterize the propagation of information in these systems. 
\end{abstract}


\maketitle
\section{Introduction}


  Chaotic behavior is usually associated with a rapid loss of information about the initial state of
  a system. In quantum systems, this can for example be quantified by studying the time dependence
  of measures for quantum information, most notably for the entanglement of two subsystems.
  Typically, chaotic systems will quickly entangle the susbsystems over time, even if they are
  initially in a product state and the spread of entanglement is usually faster than the transport
  of particles. The notion of quantum chaos is now usually connected to an effective random matrix
  theory, which is argued to be responsible for the mechanism of
  thermalization\cite{borgonovi_quantum_2016,dalessio_quantum_2016,luitz_ergodic_2016}.

The dynamical process of thermalization can be studied by a quench, where the initial
  state is prepared for example as the groundstate of a local Hamiltonian $H_0$ and the Hamiltonian
  is suddenly changed to another Hamiltonian $H$  of interest, governing the time evolution of the wave 
function. Thermalization can then be monitored by various quantities, among which
  the entanglement entropy (EE) provides a particularly appealing measure, since it encodes the
  scrambling of information about the initial state. In generic quantum systems, it grows very fast
  (a power law\cite{luitz_extended_2016}, except for the logarithmic growth in many-body localized (MBL) systems
  \cite{chiara_entanglement_2006,znidaric_many-body_2008,bardarson_unbounded_2012,serbyn_universal_2013,luitz_extended_2016}) until it saturates to a large value which scales as the volume of
  the system and is determined by the initial state (which itself is usually constructed to have low
  entanglement).

  While this scenario is very well studied, it seems clear that the scrambling of information about
  the initial state is not a property of the wave function, but rather that of the Hamiltonian. This is
  particularly plausible when thinking of two extreme cases, a generic system with diffusive
  transport, which exhibits a ballistic growth of the quench
  EE\cite{kim_ballistic_2013,luitz_extended_2016}, compared to an MBL system,
  where the growth of the quench EE is logarithmic in time and thus very slow, \emph{while using the
  same initial product state in both cases}. A multitude of previous works has established the
  differences of these two classes of systems regarding aspects of the eigenvalues and eigenvectors
  of the Hamiltonian, exhibiting \eg volume- vs area-law entanglement
  entropy\cite{bauer_area_2013,kjall_many-body_2014,vosk_theory_2015,potter_universal_2015,chen_many-body_2015,luitz_long_2016,yu_bimodal_2016} and the validity or
  violation\cite{pal_many-body_2010,luitz_long_2016,luitz_anomalous_2016} of the eigenstate thermalization
  hypothesis\cite{deutsch_quantum_1991,srednicki_chaos_1994,rigol_thermalization_2008},
  all revealing properites of the Hamiltonian.

  Motivated by the success of the study of the quench EE, we propose here to study 
  the operator entanglement entropy (opEE)\cite{bandyopadhyay_entangling_2005, prosen_operator_2007,pizorn_operator_2009,prosen_chaos_2007,musz_unitary_2013} of the
  \emph{time evolution operator}, 
which is one of those state independent measures\cite{zanardi_entangling_2000, nielsen_quantum_2003} for a quantum operation(in some references\cite{zyczkowski_duality_2004,musz_unitary_2013} termed as Shannon entropy of the reshuffled matrix). It is probably simplest to be described in the matrix product
  operator (MPO) language\cite{schollwoeck_density-matrix_2011}. 
\begin{figure}[h]
\centering
\includegraphics[width=0.4\textwidth]{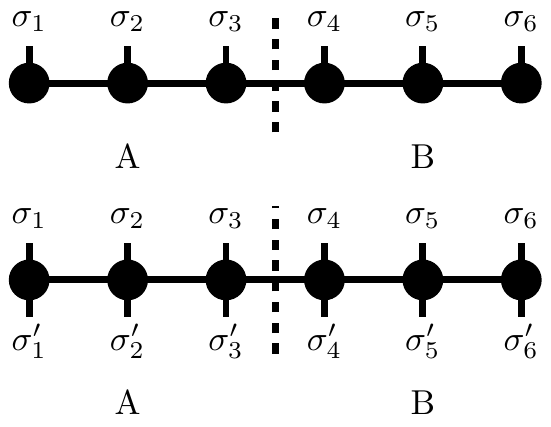}
\caption{Matrix product state (MPS) and matrix product operator (MPO). Upper panel: diagrammatic
representation of matrix product state. Each tip of the vertical bond is a physical index;
horizontal bonds contract auxillary matrix indices. After performing a left and right
canonicalization up to one bond, the chosen bond will contain all the Schmidt eigenvalues that
determine the entanglement entropy. Bottom panel: the matrix product operator can be viewed as a
matrix product state with two copies of physical indices on each site, then its entanglement entropy
can be similarly defined and calculated by a matrix product algorithm. }
\label{MPO}
\end{figure}
 As shown in Fig.~\ref{MPO}, an operator can be viewed as a matrix product state with two copies of physical indices. Then its entanglement entropy can be similarly defined as for the state. Since the time evolution
  operator of a local Hamiltonian clearly contains all the information for all possible initial states
  and correlates distant parts of the system increasingly with time, we
  expect that its opEE will grow with time and possibly saturate close to its maximal value.

  We note that the operator entanglement entropy is the relevant quantity for the efficiency of
  encoding operators as an MPO\cite{znidaric_many-body_2008,pizorn_operator_2009}, since it governs the required bond dimension. In the context of MBL,
  it was shown that the Hamiltonian is diagonalized by a unitary operator that can be represented by
  an MPO with finite bond dimension and therefore low
  opEE\cite{pekker_encoding_2014,pollmann_efficient_2016}, compared to the case of a chaotic system,
  where we expect the opEE of the diagonalizing operator to be a volume law.

  In this paper, we present the concept of the opEE of the time evolution operator and demonstrate
  that it grows as a power law in time for generic quantum systems up to a saturation value which
  scales as the volume of the system. In Sec.~\ref{sec:def}, we define the opEE of time evolution
  operator. In Sec.~\ref{sec:gen_behavior}, we map the opEE to the state EE in a quench problem as a way to understand the general growth and saturation behaviors. Then in Sec.~\ref{sec:models} we introduce the spin chain models and discuss in detail the growth and saturation of their opEE in Sec.~\ref{sec:growth} and \ref{sec:sat_val}. Finally we conclude in Sec.~\ref{sec:conclusion}. App.~\ref{app:channel_state} introduces the channel-state duality to understand opEE. App.~\ref{app:aver_rand_op_EE} contains a technical calculation of average opEE of random unitary operators. App.~\ref{app:lin_table} contains the numerical technique we used for a conserved sector of the Heisenberg spin chain.




\section{Operator Entanglement Entropy (opEE) of the Time Evolution Operator}
\label{sec:def}
\subsection{Definition}

We begin by reviewing the definition of entanglement entropy for a pure state. Then by assigning a
Hilbert space structure for linear operators, we show that the time evolution operator is a
normalized ``wave function'' in the operator space and therefore the entanglement entropy for it can be naturally defined.  

\subsubsection{Wave function entanglement entropy}

For a real space bipartition into subsystems
$A$ and $B$, a pure state $\ket{\psi}$ in the Hilbert space $\mathcal{H}$ can be represented in a basis which is the
tensor product of the orthonormal bases of the two subsystems $\{| i\rangle_A\}$ and $\{|j
\rangle_B\}$:
\begin{equation}
|\psi \rangle = \sum_{ij} \psi_{ij} |i\rangle_A \otimes  |j \rangle_B
\end{equation}
where the coefficients are given by the inner product
\begin{equation}
\psi_{ij} = \Big( {}_A\langle i | \otimes {}_B \langle j | \Big) \, | \psi \rangle.
\end{equation}
The reduced density matrix of subsystem $A$ is then obtained by performing the partial trace of the
pure density matrix $\ket{\psi}\langle\psi|$ over the subsystem $B$:
\begin{equation}
\rho_A = \tr_B \ket{\psi}\bra{\psi} \quad \Rightarrow \quad \rho^A_{ij} = \sum_k \psi_{ik} \psi_{jk} ^*
\end{equation}
and the von Neumann entanglement entropy (which corresponds to the R\'enyi entropy with
R\'enyi index $q=1$) is the Shannon entropy of its eigenvalues $\lambda_n$:
\begin{equation}
S_A =  - \tr( \rho_A \ln \rho_A  ) = -\sum_n \lambda_n \ln \lambda_n.
\end{equation}

\subsubsection{Operator entanglement entropy}

The concept of the entanglement entropy introduced above has been generalized to the space of operators\cite{prosen_operator_2007}, which is also a Hilbert space $\{ \frak{O}, (\cdot,\cdot) \}$ with the inner product $(\cdot,\cdot) :
\frak{O}\times\frak{O}\rightarrow \mathbb{C}$ on the space of linear operators $\frak{O}$ on
$\mathcal{H}$. The inner product for two operators $\hat O_1, \hat O_2 \in \frak{O}$ is defined by

\begin{equation}
    ( \hat O_1, \hat O_2 ) = \frac{1}{\sqrt{\text{dim}({\frak{O}})}}\tr( \hat O_1^{\dagger} \hat O_2 ).
    \label{eq:inner-prod}
\end{equation}
Here $\text{dim}({\frak{O}})=\text{dim}(\mathcal{H})^2$ is the dimension of the operator space. 
We can now construct two complete basis sets $\{\hat A_i\}$ and $\{\hat B_i\}$ which are orthonormal with respect to
the inner product \eqref{eq:inner-prod} and consist only of operators with a support on
the subsystems $A$ and $B$ respectively. The two bases span operator spaces $\frak{O}_A =
\text{span}(\{\hat A_i\})$ on the subsystem $A$ (and $B$ respectively) and their tensor product is
the full operator space $\frak{O} = \frak{O}_A \otimes \frak{O}_B$.

For any linear operator $\hat O \in \frak{O}$, we then have a unique decomposition
\begin{equation}
    \hat O = \sum_{ij} O_{ij} \hat A_i \otimes \hat B_j 
\end{equation}
with coefficients $O_{ij}\in \mathbb{C}$ obtained by the inner product
\begin{equation}
O_{ij} =  ( \hat A_i \otimes \hat B_j , \hat O ).
\end{equation}

In particular, we consider the unitary evolution operator
\begin{equation}
 \hat{U}(t) = \mathcal{T} e^{ - i \int_0^t H(t') dt' }
\end{equation}
given in general by a time ordered exponential. It propagates the wave function from time zero to
time $t$ and satisfies the unitary condition at all times $t$
\begin{equation}
    \Big(\hat{U}(t),\hat{U}(t)\Big) = \frac{1}{\text{dim}(\mathcal{H})}\tr\left[ \hat U(t)^{\dagger}
    \hat U(t) \right] = 1.
\end{equation}
As a result, it is a normalized element of the operator space $\frak{O}$ in the same way as a normalized
wave function in the Hilbert space $\mathcal{H}$.

With these ingredients, we can define the \emph{operator entanglement entropy} (opEE) as the Shannon
entropy of the eigenvalues of the reduced \emph{operator density matrix}
\begin{equation}
S = - \tr( \rho^A_{\rm op}  \ln \rho^A_{\rm op}  ) 
\end{equation}
where the operator reduced density matrix in this basis is
\begin{equation}
(\rho^A_{\rm op })_{ij}  = \sum_k U_{ik} (U^{\dagger})_{kj},
\end{equation}
with $U_{ij} = (\hat A_i \otimes \hat B_j, \hat U(t) )$.

When $t = 0$, the evolution operator is the identity operator and hence has zero initial opEE. As
$t$ increases, the operator becomes more and more complicated and we expect the opEE to reflect the
complexity of the time evolution. To ease the notation, we will drop the hat if its operator nature
is clear in the context, but  will reserve it in figures to make the difference to the usual
``state'' EE explicit.




\section{General Behavior of opEE}
\label{sec:gen_behavior}
\subsection{Mapping to Quench}
\label{sec:map_quench}

It is useful to map the opEE of the time evolution operator to a quench problem in a larger
Hilbert space, since this allows us to connect to known features of
the wave function entanglement entropy. In App.~\ref{app:channel_state}, we introduce one possible
mapping via the channel-state duality, but this is not the only possibility(for example see \onlinecite{wang_entangling_2003} for one using swap operation and \onlinecite{prosen_operator_2007} for another map in Majorana representation, etc.). Let us introduce below
a different mapping preserving locality, which makes it easier to identify the general features of the opEE of the time
evolution operator as a function of time in finite systems. Specifically, for an evolution operator
(of a time independent Hamiltonian) $U(t) =e^{ - i Ht} $, our goal is to construct a corresponding $\overline{H}$ and state $|\psi \rangle$, such that state EE of $e^{-i \overline{H}t } | \psi\rangle $ is the same as opEE of $U(t)$ under the same real space partition:
\begin{equation}
U(t)  = e^{-iHt}\quad  \leftrightarrow  \quad e^{ - i \overline{H} t} | \psi \rangle. 
\end{equation}

We will see in the next two subsections that in the MBL phase, a simple choice is
\begin{equation}
\overline{H} = H, \quad |\psi \rangle = \otimes_{i=1}^{L} |\!\!\uparrow \rangle 
\end{equation}
where $L$ is the number of sites. Whereas for a generic Hamiltonian, $|\psi\rangle$ can be taken as
\begin{equation}
|\psi \rangle = \otimes_{i=1}^{2L} | \!\!\uparrow \rangle 
\end{equation}
which is in a two-copy Hilbert space $\mathcal{H} \otimes \mathcal{H}$ that can represent all possible operators in $\frak{O}$. $\overline{H}$ in this case will be an operator acting on $\mathcal{H} \otimes \mathcal{H}$, which we will construct explicitly below.

\subsubsection{MBL Hamiltonian}

In the MBL phase the Hamiltonian is effectively given by multi-spin interaction terms between the
``l-bits''
$\sigma^x_i$\cite{serbyn_local_2013,serbyn_universal_2013,huse_phenomenology_2014,nandkishore_many-body_2015,imbrie_many-body_2016,imbrie_diagonalization_2016,imbrie_review:_2016} 
\begin{equation}
H_{\rm MBL} = \sum_{i} h^i_0 \sigma_i^x + \sum_{ij} J_{ij}^0 \sigma_i^x \sigma_j^x  + \sum_{ijk} J_{ijk}^0 \sigma_i^x  \sigma_j^x \sigma_k^x + \cdots 
\end{equation}
The ``l-bits'' are local integrals of motion, and the exponentially (with distance) decaying
interactions between them are responsible for the slow entanglement dynamics in MBL
systems\cite{chiara_entanglement_2006,znidaric_many-body_2008,serbyn_universal_2013,bardarson_unbounded_2012,luitz_extended_2016,singh_signatures_2016}.
For this particular type of Hamiltonian, the time
evolution operator $U(t)$ will only consist of products of the identity operator $I\equiv \sigma^0$ and $\sigma^x$.
In contrast, the local basis of the complete operator space $\frak{O}$ consists of the 4 elements
$I_i$, $\sigma^x_i$, $\sigma^y_i$ and $\sigma_i^z$ on site $i$ of the system(compactly denoted as $\sigma_i^{\mu}$ with $\mu = 0,1,2,3$ for later convenience).
This means that the MBL time evolution operator is only contained in a subspace of dimension $2^L$
of the total operator Hilbert space $\frak{O}$ which has the dimension $\dim\left(\frak O\right) =
4^L$. This allows us to map the evolution operator to a state Hilbert space of dimension $2^L$
\emph{without doubling the number of degrees of freedom}. 

Note that the states $\ket{\!\!\up} = I\ket{\!\!\up}$ and $\ket{\!\!\dw} = \sigma^x |\! \up \rangle$
are orthonormal in the single-site Hilbert space and we can therefore map $I\rightarrow
I\ket{\!\!\up}$ and $\sigma^x \rightarrow \sigma^x \ket{\!\!\dw}$. In the multiple-site situation, the basis which only consists of products of $I\equiv \sigma^0$ and $\sigma^x$
\begin{equation}
\sigma_{1}^{\mu_1} \sigma_{1}^{\mu_2}  \cdots \sigma_{L }^{\mu_L}  \quad \mu_i = 0 \text{ or } 1 
\end{equation}
can be mapped to orthonormal basis in $\mathcal{H}$ as 
\begin{equation}
\sigma_{1}^{\mu_1} \sigma_{1}^{\mu_2} \cdots \sigma_{L }^{\mu_L} |\!\up \cdots \up \rangle. 
\end{equation}

Then the decomposition of $ U(t) | \!\up \dots \up\rangle$  in the state basis is identical to the
decomposition of $U(t)$ in the operator basis and the state EE of the wave function after a global
quench from the state $|\!\!\up \cdots \up \rangle$ given by $\ket{\psi(t)} = U(t) |\!\up \cdots \up
\rangle$ is identical to the opEE. Therefore in this case the bar transformation is the trivial identity map and $|\psi \rangle = \otimes_{i=1}^L | \!\! \up \rangle $.

Applying well known results on the logarithmic EE growth after a quench from a product state in MBL
systems\cite{chiara_entanglement_2006,znidaric_many-body_2008,bardarson_unbounded_2012,serbyn_universal_2013,luitz_extended_2016,singh_signatures_2016}, this mapping immediately implies a logarithmic long time growth and
an extensive submaximal saturation value of the opEE in MBL systems and therefore gives an initial state independent description of the information propagation in MBL. 

\subsubsection{Generic Hamiltonian}

A completely generic (and possibly non-local) Hamiltonian can be composed of all possible spin interactions
\begin{equation}
H = \sum_{\mu_1, \mu_2, \cdots, \mu_L = 0 }^3 J_{\mu_1 \mu_2 \cdots \mu_L }  \sigma_1^{\mu_1}\sigma_2^{\mu_1} \cdots \sigma_L^{\mu_1},
\end{equation}
where $\sigma^0\equiv I$ is understood as identity operator and $J_{\mu_1 \mu_2 \cdots \mu_L }$ are complex interaction coefficients (this is in fact an operator basis decomposition of the Hamiltonian). It occupies the full operator Hilbert space $\frak{O}$, and so by dimensional counting, it is only possible to map the operator to a double-site state Hilbert space $\mathcal{H}\times \mathcal{H}$. In order to do so, we equip each site with an auxiliary site and upgrade the Hamiltonian $H$ to 
\begin{equation}
\overline{H}  = \sum_{\mu_1, \mu_2, \cdots, \mu_L = 0 }^3 J_{\mu_1 \mu_2 \cdots \mu_L }  \overline{\sigma}_1^{\mu_1} \overline{\sigma}_2^{\mu_1} \cdots \overline{\sigma}_L^{\mu_1},
\end{equation}
where each $\overline{\sigma}$ is acting on both the physical site and the nearby auxiliary site. For one dimensional systems, it is helpful to think of the new Hamiltonian $\overline H$ as a ladder system
(bilayer for 2d), which implies that the locality features of the initial system are preserved.
The ``bar'' transformation is defined by the mapping
\begin{equation}
    \begin{split}
\overline{I} &=  I \otimes I, \quad\quad \overline{\sigma}^x =  \sigma^x \otimes \sigma^x,\\
\overline{\sigma}^y &= \sigma^y \otimes I, \quad \overline{\sigma}^z = \sigma^z \otimes \sigma^x,
\end{split}
\end{equation}
where we introduce a $\sigma^x$ operator on auxiliary degrees of freedom for every operator except for the
identity (producing an identity on the auxiliary site).

The bar transformation is chosen such that 
\begin{equation}
\{\overline{\sigma}^{\mu} | \uparrow \uparrow\rangle\} 
\end{equation}
corresponds to an orthonormal basis of the local Hilbert space corresponding to one site and its
auxiliary site. Therefore by the same generalization to multiple sites, the opEE of $U(t)$ will be
identical to the wave function entanglement entropy of $\ket{\psi(t)} = \overline{U}(t) | \uparrow \cdots \uparrow \rangle$. 

On the other hand, the bar transformation is an operator algebra isomorphism, which means for operators $O_1, O_2\in \frak{O}$, we have
\begin{equation}
\overline{O_1 O_2} = \overline{O}_1 \overline{O}_2.
\end{equation}
For the time evolution operator of a time independent Hamiltonian (or each infinitesimal time evolution in the time ordered product), we have
\begin{equation}
\begin{aligned}
\overline{ e^{-i Ht } } &= \overline{ \lim_{n\rightarrow \infty} ( 1 - i H \frac{t}{n})^{n} } \\
&= \lim_{n\rightarrow \infty} \overline{(1 - i H \frac{t}{n})^n}  = \lim_{n\rightarrow \infty} (1 - i \overline{H} \frac{t}{n})^n \\
&= \exp( - i \overline{H} t ) .
\end{aligned}
\end{equation}
As a result, the opEE of $U(t)$ is equal to the state EE of $\exp( - i \overline{H} t ) |\uparrow \cdots \uparrow \rangle$. 

As argued above, the barred Hamiltonian will preserve the locality of interaction, (non-)integrability of the model and the spatial disorder distributions across the system. As a result, our knowledge of the global quench EE of those systems can be carried over. 

We therefore expect that the opEE of the evolution operator will in general have three domains in
its evolution with time. The behavior for times less than propagating over one lattice spacing
should not be taken seriously, because of its regulator dependence. In the intermediate region when
the EE is propagating through the system, the opEE will increase in a certain manner that is described
by some scaling function (e.g. linear growth in CFT\cite{calabrese_evolution_2005}, power law in
thermal phase\cite{kim_ballistic_2013,luitz_extended_2016}, logarithmic growth in MBL
phase\cite{chiara_entanglement_2006,znidaric_many-body_2008,bardarson_unbounded_2012,serbyn_universal_2013,luitz_extended_2016,singh_signatures_2016}). In the thermal phase, the opEE will reach a saturation value which is
extensive, however for integrable systems this may not be true due to the possible recurrence
behavior\cite{cardy_quantum_2016}.



\subsection{The Page Value}

Before analyzing the unitary time evolution of a physical Hamiltonian, let us first consider the average opEE of a random unitary operator, which will give us a guideline for the saturation behavior of random systems. 

The average EE of a random wave function (with a measure that is invariant under unitary transformations,
i.e. Haar measure) was derived by Page\cite{page_average_1993} to be
\begin{equation}
\begin{aligned}
S_{\rm Page}  &= \left( \sum_{k=n+1}^{mn} \frac{1}{k }\right) - \frac{m - 1}{2n} \\
&= \ln m - \frac{m}{2n} + \mathcal{O}(\frac{1}{mn}),
\end{aligned}
\end{equation}
where $m$ and $n$ are the dimensions of the Hilbert space of the two subsystems and $m \le n$. 

To fix the notation for clarity, let us consider $L$ sites hosted with $j = \frac{1}{2}$ spins in a chain with a bipartition
in a smaller system (A) composed of $\ell_A$ spins (its complement B has then $\ell_B = L - \ell_A$ spins), then
\begin{equation}
S_{\rm Page}[\psi] = \ell_A \ln 2 - 2^{\ell_A - \ell_B - 1 } + \mathcal{O}(2^{-L}).
\end{equation}
For an equal partition ($\ell_A = \ell_B$)
\begin{equation}
  S_{\rm Page}[\psi]  = \frac{L}{2} \ln 2  - \frac{1}{2} + \mathcal{O}(2^{-L}).
\end{equation}
The deficit $\frac{1}{2}$ from the maximal possible EE here suggests a deviation from a maximally
entangled state on average.

In App.~\ref{app:aver_rand_op_EE}, we perform an analytic calculation based on an integration technique on the unitary group to show that the average opEE of random gate(unitary operator) is given by the Page value of a doubled system
\begin{equation}
S_{\rm Page} [\hat{U}] =  2\ell_A \ln 2 - 2^{2\ell_A - 2\ell_B - 1 } + \mathcal{O}(4^{-L}). 
\end{equation}
The deficit for the even partition is again $\frac{1}{2}$. 

In the analysis of our numerical results for the opEE of the evolution operator, we will compare the
saturation values to the Page value.




\section{Models}
\label{sec:models}
\begin{figure}[h]
    \centering
    \includegraphics{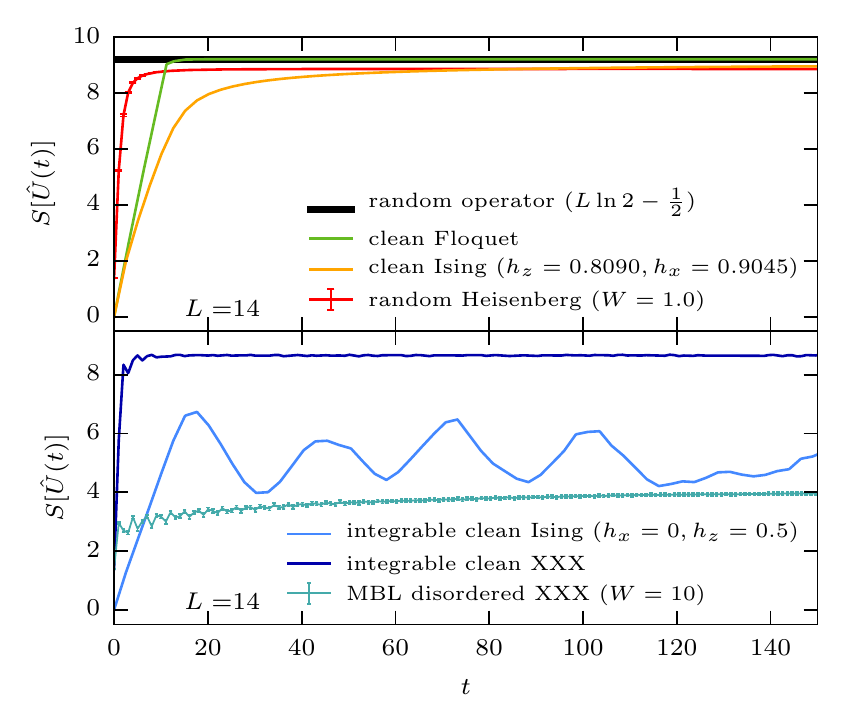}
    \caption{Operator entanglement entropy $S[\hat{U}(t)]$ of the time evolution operator
    $\hat{U}(t)$ for the models introduced in Sec.~\ref{sec:models} as a function of time $t$. Top
    panel: Chaotic models. Bottom panel: Integrable (MBL) models. For all models, the opEE grows fast
for short times and for the considered nonintegrable models, the opEE reaches a large value close to
that of a random operator ($L \ln 2 -\frac 1 2$). For the clean integrable cases, the opEE
fluctuates at large times due to commensurate periods of integrals of motion. The MBL model exhibits the
slowest growth of the opEE with time. For the Floquet model, the opEE grows nearly linearly and
saturates at the limit of a random unitary operator. The results for disordered models are averaged
over 100 to 1000 realizations.
}
    \label{fig:opEEoverview}
\end{figure}

   We study various spin models to investigate the behavior of the opEE as a function of time. One
    of the simplest models is the Ising model in a tilted field
    \begin{equation}
        H = \sum_i \sigma^z_i \sigma^z_{i+1} + h^x_{i} \sigma^x_i + h^z_{i} \sigma^z_i,
        \label{eq:Ising}
    \end{equation}
with the Pauli matrix $\sigma^x_i$ and $\sigma_i^z$ on site $i$. Here, $h^x_{i}$ ($h^z_{i}$) are the
transverse (longitudinal) magnetic fields and we choose a homogeneous configuration without disorder. This model is
integrable in the clean case if either $h^x$ or $h^z$ vanish, and non-integrable if both fields are
non-zero with generic parameter choices. We adopt a popular set
\begin{equation}
h^x = 0.9045\quad h^z = 0.8090
\end{equation}
to compare it with existing literature where it was argued that the system becomes robustly
nonintegrable with these parameters\cite{prosen_general_2002,kim_ballistic_2013,zhang_thermalization_2015,zhang_floquet_2016,akila_particle-time_2016-1,akila_particle-time_2016}.

    We also study the \emph{standard model} of MBL, the random field Heisenberg chain 
    \begin{equation}
    H = \frac 1 4 \sum_i \left[ \sigma_i^x \sigma_{i+1}^x + \sigma_i^y \sigma_{i+1}^y + \sigma_i^z
    \sigma_{i+1}^z \right] + \frac{h_i^z}{2} \sigma_i^z 
        \label{eq:Heisenberg}
    \end{equation}
    where $h_i$ is the random field on site $i$. We take it to be drawn from a uniform distribution on the interval $[-W,W]$, where $W$ is the disorder strength. This model has an MBL transition at a disorder strength of $W\approx
    3.7$ \cite{pal_many-body_2010,luitz_many-body_2015} and recent numerical evidence points to slow dynamics and subdiffusion at
    weaker disorder
    \cite{bar_lev_dynamics_2014,bar_lev_absence_2015,agarwal_anomalous_2015,luitz_extended_2016,znidaric_diffusive_2016,luitz_ergodic_2016}. It conserves the total
    magnetization $S_z=\sum_i S_i^z$, and allows therefore to study only one magnetization sector.
    We project to the largest sector with $S_z=0$, having a Hilbert space dimension of
    $\binom{L}{L/2}$. In the clean case $W=0$, it is the integrable XXX chain.

    While the Ising model in a tilted random field does not conserve magnetization, it still
    conserves energy. In order to have a completely generic quantum system, one can even break
    energy conservation by introducing periodic driving. We study a Floquet system with driving
    period $\tau$ given by the time evolution operator over one period
\begin{equation}
\label{eq:floq}
U(\tau) = \E^{ - \I 2H_{xy} \frac{\tau}{2} }  \E^{ -\I 2 H_z \frac{\tau}{2} }
\end{equation}
where
\begin{equation}
\begin{aligned}
H_{xy} &= \sum_i h_x^i \sigma^x_i + h_y^i \sigma^y_i , \quad h_x^i = 0.9045\\
H_z &= \sum_i \sigma^z_i \sigma^z_{i+1} +  h^i_{z} \sigma^z_i, \quad h_z^i = 0.8090
\end{aligned}
\end{equation}
We set $\tau = 0.8$ and make two choices of $h_y^i$. 
\begin{enumerate}
\item $h_y^i =0$. This case has the same time averaged Hamiltonian as the chaotic Ising model
    evolution\cite{zhang_floquet_2016}. However the system is time-reversal invariant (by shifting
    half of the period, the unitary matrix is symmetric) and therefore corresponds to circular orthogonal ensemble (COE).
\item $h_y^i = 0.3457$. The $\sigma^y$ term breaks the time reversal symmetry and thus $U(\tau)$
    should be in the circular unitary ensemble (CUE).
\end{enumerate}

Both choices lead to essentially identical behavior of the opEE, although in the CUE case, the
dynamics seems to be slightly more chaotic, as we discuss in Fig. \ref{fig:floqgrowth}. With this
exception, we study the COE model in the rest of this paper. We employ open boundary conditions throughout this work.  

Fig.~\ref{fig:opEEoverview} gives an overview of the results for all these models with system size $L = 14$. In accordance with our state-quench mapping, the opEE has a fast growth at short times (except for
the MBL case) and then saturates to a constant value for thermal phase, possibly oscillating in the integrable models. In the next few sections, we will address in detail the saturation value and the scaling of the growth respectively.




\section{Saturation Value}
\label{sec:sat_val}

Let us first focus on the long time behavior of opEE $S[\hat{U}(t)]$ for various models. In nearly all systems
that we considered, the opEE saturates to a constant value in Fig.~\ref{fig:opEEoverview} at
sufficiently long times, and we classify them as follows: Maximally scrambling behavior is found in
the Floquet system, where $S[\hat U(t)]$ saturates to the Page value corresponding to a random
unitary operator sampled from the Haar measure. Chaotic systems with conservation laws (energy,
magnetization) also exhibit a large saturation value close to the Page value but with a
small deficit which seems to be independent of system size and more conservation laws seem to lead
to a larger deficit. In the MBL system with local integrals of motion the opEE saturates after a very long time
at a value much smaller than the Page value, whereas clean integrable models with nonlocal
conservation laws show no saturation of the opEE, but rather fluctuate more or less strongly around a
value that is smaller than the Page value. From this observation, we speculate that the specific
nature of the integrals of motion and presumably their incompatibility with the real space partition
causes these fluctuations.

We will devote the rest of this section to the details of the two chaotic classes.




\subsection{Floquet Spin Model}
\label{sec:floq_sat}
We study the Floquet spin model \eqref{eq:floq} introduced in Sec.~\ref{sec:models} as a typical model with no
conservation laws. In previous studies, the average global quench EE of an initial product state
after a long time evolution with the Floquet Hamiltonian was found to be given by the Page value $\frac{L}{2}\ln 2 -\frac 1 2$\cite{zhang_thermalization_2015,zhang_floquet_2016}.

Fig.~\ref{fig:opEEoverview} illustrates that the numerically calculated opEE for the equal bipartition ($\ell_A = \ell_B$) saturates to
$L\ln 2 -\frac 1 2$ in the long time limit. In App.~\ref{app:aver_rand_op_EE}, we show that this
is the average opEE of a random unitary operator by partly using Page's result for a
random state\cite{page_average_1993}. This saturation value is in agreement with the
consensus\cite{haake_quantum_2010} that the Floquet evolution operator (without time reversal symmetry) is indeed a physical realization of the circular unitary ensemble. 

\begin{figure}[h]
\centering
\includegraphics[width=\columnwidth]{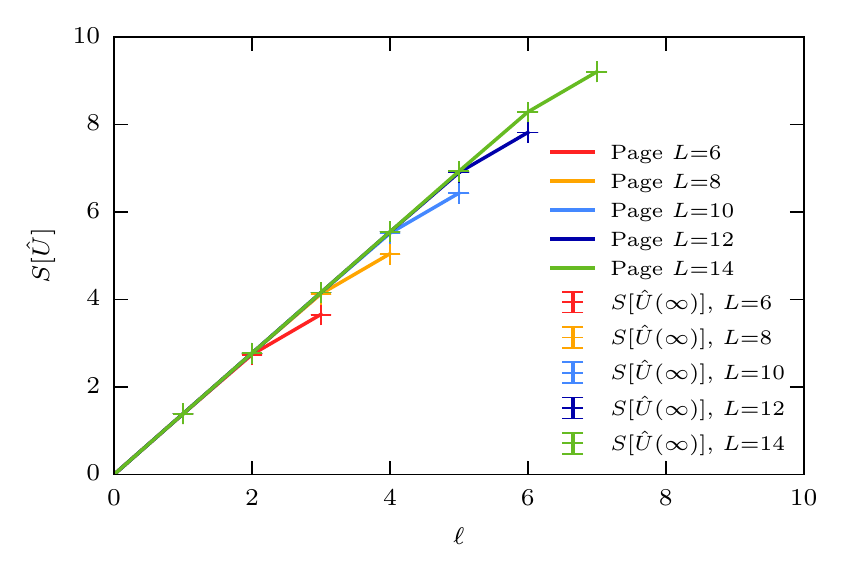}
\caption{Comparison of the saturation value of the opEE of the time evolution operator $S[\hat
U(\infty)]$ (crosses, errorbars illustrate the size of fluctuations around the saturation value) of
the Floquet model with the result for a random unitary operator (Page value, given by full lines)
for all possible smaller subsystem sizes $\ell$. The opEE for the Floquet system matches the Page
result for all system sizes and partitions perfectly.}
\label{fig:page_curve}
\end{figure}

In order to confirm this, we calculate the long time opEE $S[\hat U(\infty)]$ of the Floquet evolution operator for all possible bipartitions of the system and compare the results in Fig.~\ref{fig:page_curve} to 
\begin{equation}
\label{eq:op-page-val}
S_{\rm Page} = 2 \ell_A \ln 2 - 2^{2\ell_A - 2\ell_B - 1 } 
\end{equation}
which is essentially the average opEE of a random unitary operator in the corresponding partition,
where $\ell_A$ is the length of subsystem $A$ and $\ell_B = (L-\ell_A)$ the correspondingly the length of its
complement. The Floquet evolution operator opEE matches to the Page value for all
possible partitions even in very small systems.



\subsection{Chaotic systems with conservation laws}

  The next set of examples we consider in Fig.~\ref{fig:opEEoverview} are generically nonintegrable systems with conservation laws,
  in particular the random field Heisenberg chain at weak disorder (such that it does not exhibit
  MBL\cite{luitz_many-body_2015}) and the tilted
  Ising chain. The former conserves energy and total magnetization, while the later was shown to be generically nonintegrable in Ref.  \onlinecite{kim_ballistic_2013} and conserves energy and parity under reflection. 

We choose to present in detail the results of the random field Heisenberg model to show the
different behaviors owing to the conservation laws and the locality of interactions. 

\begin{figure}[h]
    \centering
    \includegraphics{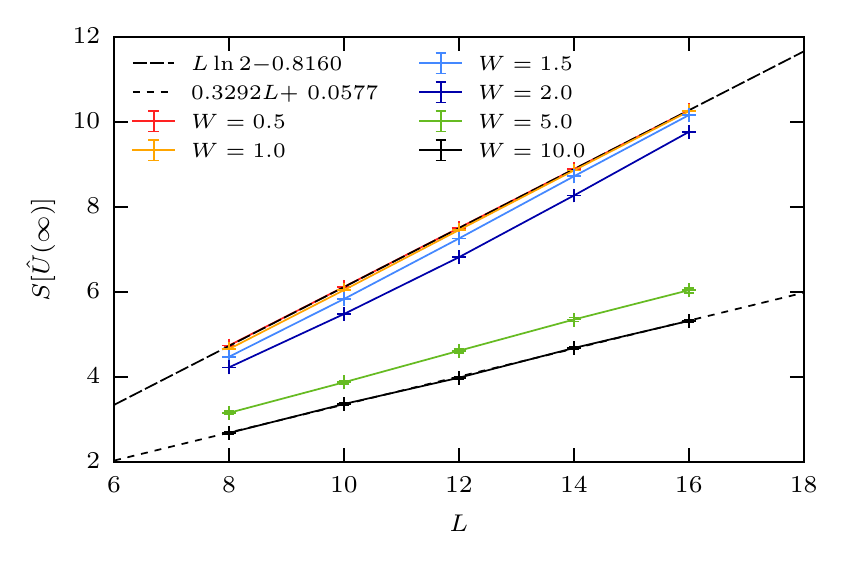}
    \caption{Saturation value of the disorder averaged opEE for the equal bipartition for the random field Heisenberg model as a function of
    system size for different disorder strengths. For strong disorder $W\gtrsim 3.7$, the system is
in the MBL phase and we observe a suppressed but extensive saturation value. At weak disorder, the
saturation value scales as $L\ln 2$ but has a constant deficit.}
    \label{fig:opEE_saturation}
\end{figure}

The saturation behaviors for various disorder strengths and different system sizes are shown in
Fig.~\ref{fig:opEE_saturation}. For weak disorders, the saturation values is still around $L \ln 2$,
but the deficit value is larger than $\frac{1}{2}$. When the disorder strength is so large that the
system is in the MBL phase, the saturation value is extensive but is only a fraction of $L \ln 2$. Note
that at weak disorder, there are visible finite size effects, as it seems that for larger system
sizes, the saturation values for different disorder strengths approach each other. We suspect that
for much larger system sizes, the deficit for disorder strengths below the MBL critical disorder
becomes in fact equal, but will remain larger than $\frac 1 2$.

The MBL behavior is easy to be interpreted from the state-quench mapping discussed in
Sec.~\ref{sec:map_quench}. In fact, for systems deep inside MBL phase, the opEE can be directly
mapped to the global quench of the \emph{same} Hamiltonian, and so an upper bound for the saturation
value is $\frac{L}{2} \ln 2$, which is indeed far less than the value in large chaotic systems. 

In the next section, we ascribe the opEE deficit in the thermal phase to the block structure of the
reduced density matrix, which ultimately is a result of the conserved total magnetization. We
believe that a similar reasoning can also be applied to other thermal phase models with conservation
laws, but an explicit demonstration is lacking.




\subsection{Deficit Value of Random Field Heisenberg Model}

The thermal phase saturation value of the chaotic models we studied is close to the maximal value, but the deficit is
greater than that of the Floquet systems. Here we present an argument to explicitly show how the
conservation law is responsible for this fact in a fixed magnetization sector ($S_z=0$) of the random field Heisenberg model. 

For simplicity, we will present this argument for the EE of a wave function and explain the generalization to the opEE in the end of this subsection.

The constraint $S_z = 0$ tells us that the $S_z$ bases for part $A$ and $B$ have to be complementary,
\ie only states with $n_{\up}$ up spins in A and $N - n_{\up}$ up spins in B can be paired to form
the basis of $S_z = 0 $ sector of $2N$ sites (other combinations will yield a vanishing wave
function coefficient).

Thus for any given state $|\psi\rangle$ with fixed magnetization $S_z = 0$, we can do a decomposition
\begin{equation}
|\psi \rangle = \sum_{n_{\up} = 0}^N \sum_{ij} \psi_{n_{\up} }^{ij} | n_{\up}, i \rangle_A | N- n_{\up}, j \rangle_B 
\end{equation}
where $\psi^{ij}_{n_{\up}}$ is a block matrix with ${ N \choose n_{\up}}$ rows and columns and $ij$ are the row and column indices. This implies that the reduced density matrix is also block diagonal with size ${ N \choose n_{\up}}$ for each block. The operator version of
this decomposition is just the one with a two-copy block size ${ N \choose n_{\up}} { N \choose n_{\up}'}$.
App.~\ref{app:lin_table} gives a detailed account of how to utilize this structure in numerical computations. 

\begin{figure}[h]
\centering
\includegraphics[width=0.4\textwidth]{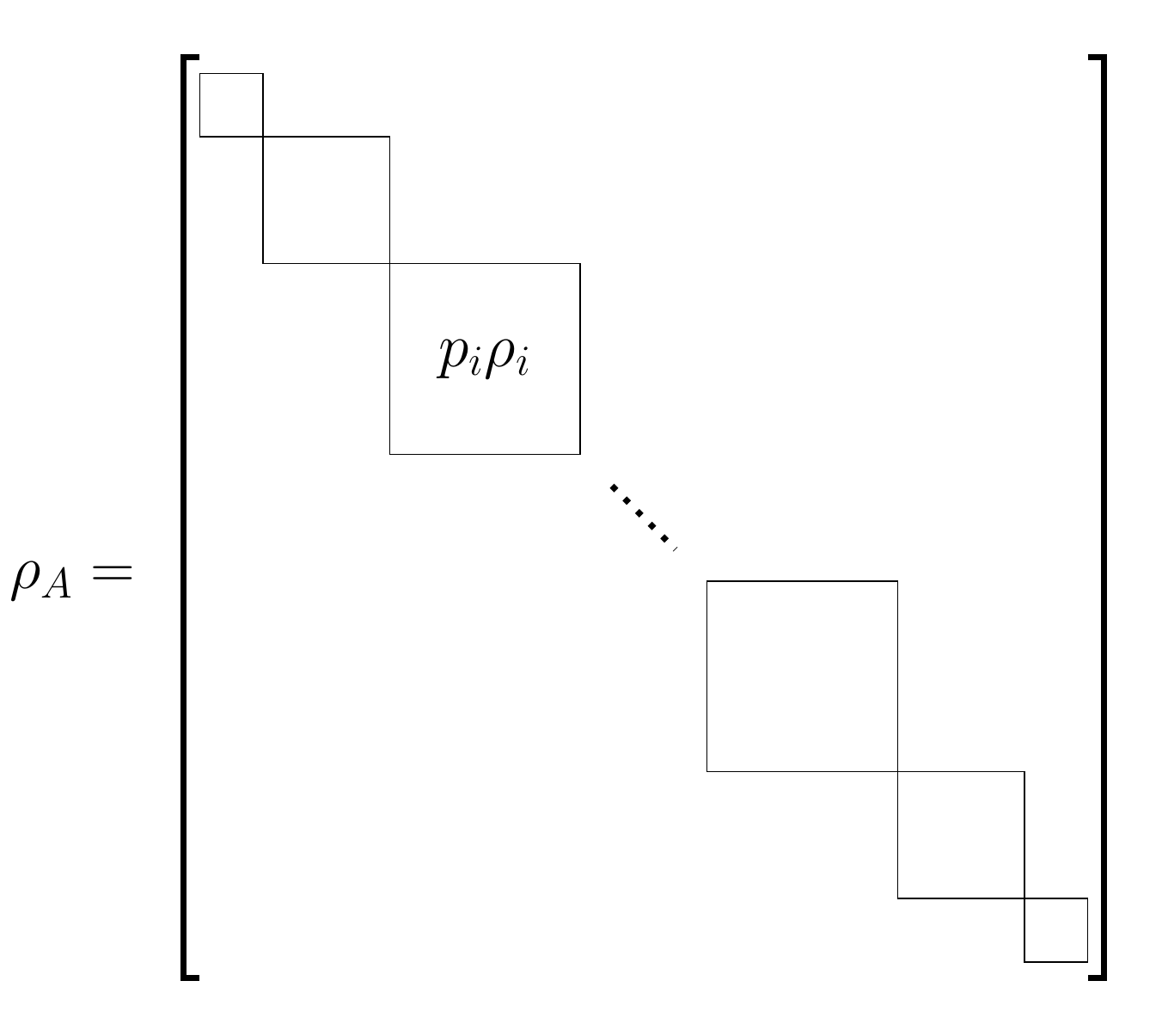}
\caption{Block diagonal form of the reduced density matrix. }
\label{block.pdf}
\end{figure}

Let the eigenvalues of the $i$-th block be $p^{(i)}_j$ (the index $i$ is the same as $n_{\up}$), and
the trace of block $i$ be $p_i = \sum_{j} p^{(i)}_j$, then the EE is 
\begin{equation}
S =  - \sum_{i, j} p^{(i) }_j  \ln p^{(i)}_j  = - \sum_{i} p_i \ln p_i + \sum_i p_i S_i ,
\end{equation}
where $S_i$ is the EE of the $i$-th block
\begin{equation}
S_i = - \sum_j \frac{p^{(i)}_j}{p_i} \ln \frac{p^{(i)}_j}{p_i}.
\end{equation}
If we regard each block rescaled by $p_i$ to be a density matrix $\rho_i$, the total density matrix is then a classical mixture
\begin{equation}
\rho = \sum_i p_i \rho_i \qquad [\rho_i, \rho_j ] = 0
\end{equation}
and the EE is the sum of the occupation entropy plus the average EE of all the blocks. 

In principle, one should average the expression of $S$ over a probability distribution of the occupation probability $p_i$ and entropy distribution of $S_i$. For the sake of obtaining an upper bound, we can avoid this complication by just calculating the maximal value. The optimization problem
\begin{equation}
\text{max}_{p_i} S[\{p_i\}] \quad\text{ subject to } \quad \sum_i p_i = 1 
\end{equation}
is easily solved by the Lagrangian multiplier technique and the maximal EE taking place at $p_i = \frac{e^{S_i}}{\sum_i e^{S_i} }$ is
\begin{equation}
S_{\rm max}  = \ln \sum_i e^{S_i}
\end{equation}

\begin{figure}[h]
    \centering
    \includegraphics[width=\columnwidth]{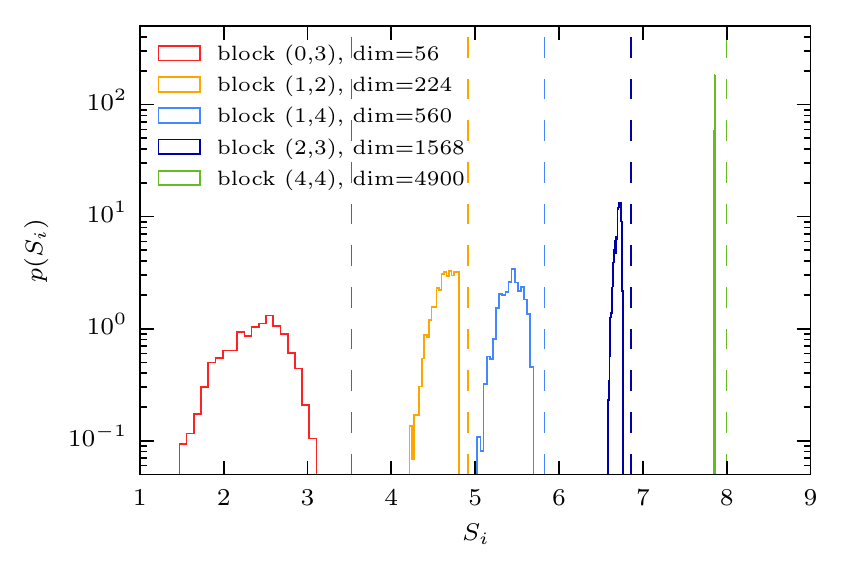}
    \caption{Probability distribution of the entropy $S_i$ of different blocks of the operator
    reduced density matrix $\rho$. We take $L = 16$ disordered Heisenberg model at $W = 0.5$. The blocks are labelled by the number of up spins on subsystem $A$ in
the first and second index of $\rho$ and the Page value for each block is indicated by a
dashed line, clearly showing that it is an upper bound for the block entropy.}
    \label{fig:block_histo}
\end{figure}

If the blocks are independent and random, $S_i$ will take the corresponding Page value of the
corresponding block size. In fact, for almost all blocks, $S_i$ will be bounded by the Page value of block size${N \choose i }$ ($N = \frac{L}{2}$)
\begin{equation}
S_i <  S^i_{\text{Page}} = \ln {N \choose i} - \frac{1}{2} + \mathcal{O}(\frac{1}{N})\quad i > 0. 
\end{equation}
This is verified numerically in Fig. \ref{fig:block_histo} for different blocks of different sizes.
\begin{figure}[h]
    \centering
    \includegraphics[width=\columnwidth]{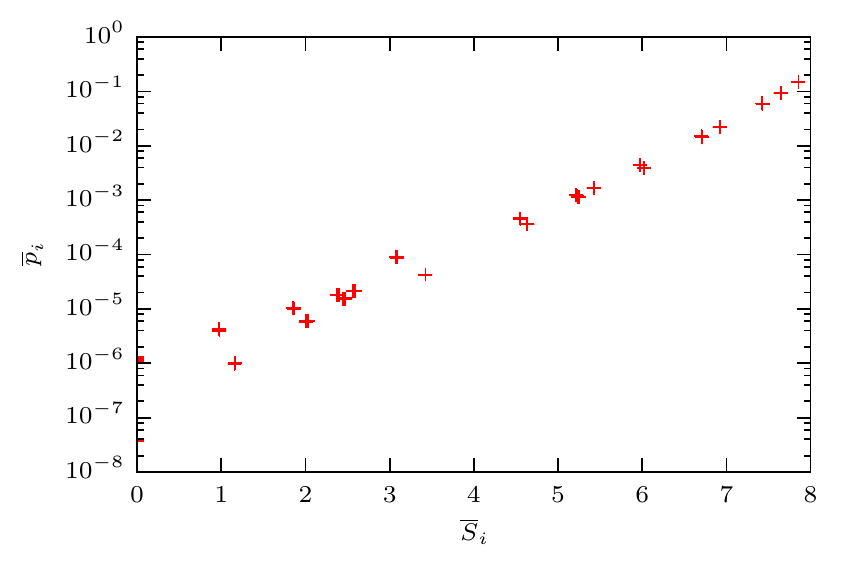}
    \caption{Average weight $p_i$ of different blocks of the operator
    reduced density matrix $\rho$ vs. the value of the block entropy $S_i$. We use the same parameters($L = 16$, $W = 0.5$) as in Fig.~\ref{fig:block_histo}.}
    \label{fig:block_weight}
\end{figure}
There are only 4 exceptional $i = 0$ blocks, which only give a $o(1)$ correction to the $\sum
e^{S_i}$, becoming an $o(\frac{1}{N})$ correction to the total entropy. Therefore the maximal value
of the opEE 
\begin{equation}
\begin{aligned}
  S_{\rm max} &< \ln \left( \sum_i e^{S^i_{\rm Page} } + o(1) \right) \\
  &= \ln \sum_i e^{S^i_{\rm Page} }  + \mathcal{O}(\frac{1}{N})\\
  &= N \ln 2 - \frac{1}{2} + \mathcal{O}(\frac{1}{N})\\
\end{aligned}
\end{equation}
is bounded by the Page value. When $N$ is large, we conclude 
\begin{equation}
S < S_{\text{max}} \le N \ln 2 - \frac{1}{2} + \mathcal{O}(\frac{1}{N} ).
\end{equation}
For the operator version, the only change is the block size which becomes ${N \choose i} {N \choose j}$, and hence
\begin{equation}
    S_\text{op} < \ln \sum_{ij} {N \choose i} {N \choose j} - \frac{1}{2} = 2N \ln 2 - \frac{1}{2}  = L \ln 2 - \frac{1}{2}.
\end{equation}

For large blocks, the numerically calculated values for $S_i$ concentrate on the average as
illustrated in Fig. \ref{fig:block_histo}. Similarly, the average values of $p_i$ obey the optimal
distribution $\frac{e^{S_i}}{\sum_i e^{S_i}}$ as shown in Fig.~\ref{fig:block_weight}. Hence the
distribution of $p_i$ gives the largest opEE it can support for a given subblock entropy $S_i$. 
We conclude from our numerical analysis that the total deficit probably stems from the deficit
observed in each block.




\section{Growth}
\label{sec:growth}

In the previous section, we studied the behavior of the opEE of the evolution operator at very long
times in finite systems and found that a saturation value is reached. Since it is clear that at the
initial time $t=0$ the opEE is zero, we will now consider how the saturation value is reached in
several example systems.



\subsection{Growth of $S[\hat U(t)]$ for the Floquet model}
\begin{figure}[h]
    \centering
    \includegraphics{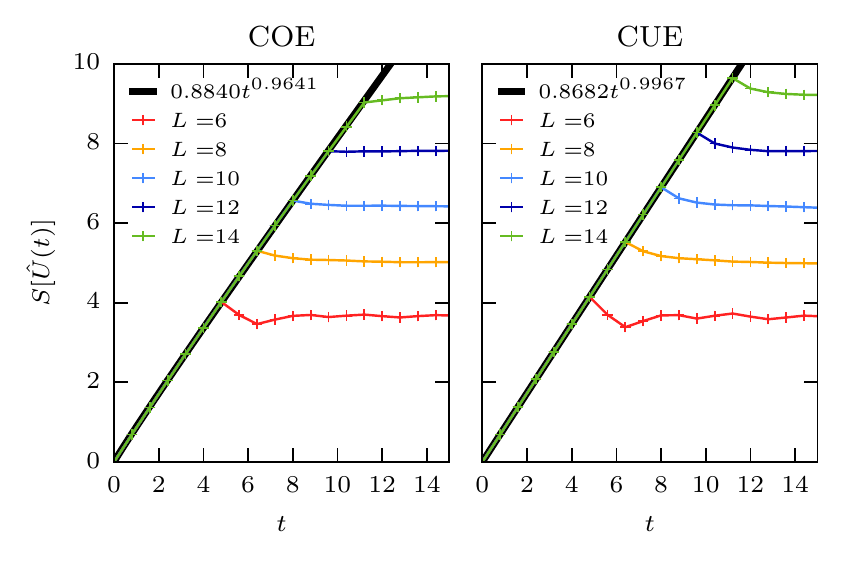}
    \caption{Growth of the operator entanglement entropy in the Floquet model \eqref{eq:floq}. The thick black lines
    correspond to a power law fit. Left panel: Floquet model with time reversal symmetry ($h_y^i =
    0$), a small deviation from a linear growth is visible in the exponent. Right panel: Floquet model without time reversal symmetry ($h_y^i = 0.3457$), the linear growth is almost perfect. }
    \label{fig:floqgrowth}
\end{figure}

In Fig.~\ref{fig:floqgrowth} we show the short time behavior of the opEE (equal bipartition) for the
Floquet model \eqref{eq:floq} of different system sizes in both COE and CUE parameter choices. Clearly, the opEE grows very fast at short
times and for different system sizes there are almost no visible finite size effects. We determine a
fit to a power law growth at short times according to the form $S[\hat U(t)] = a t^\alpha$ and
obtain an exponent of $\alpha=0.9641$ for COE system and $\alpha = 0.9967$ for CUE system. With the available system sizes it is
difficult to determine whether the discrepancy in COE system from a perfect linear growth (which is for example observed
in the growth of the quench EE in this model, starting from a product
state\cite{kim_ballistic_2013}) is a finite size effect or prevails in the thermodynamic limit, but from the robustness of the result in Fig.~\ref{fig:floqgrowth}, it is likely that the remaining time reversal symmetry leads to the deviation. After this initial almost ballistic growth, the opEE saturates to a value very close to the Page value as
discussed in Sec.~\ref{sec:floq_sat}.

These two results together are consistent with the expectation that the Floquet Hamiltonian(without conservation) can be
considered as an almost perfect scrambler. 




\subsection{Growth of $S[\hat U(t)]$ in the random field Heisenberg chain}

Let us now consider the growth of the opEE in the random field Heisenberg chain
\eqref{eq:Heisenberg}. We have already
seen that the opEE at long times saturates to a value close to the Page limit, offset by a system
size independent deficit, which we attributed to the constraints caused by conservation laws.

Here, we restrict ourselves to the case of weak disorder, where the Heisenberg chain is not fully
many-body localized.

\begin{figure}[h]
    \centering
    \includegraphics{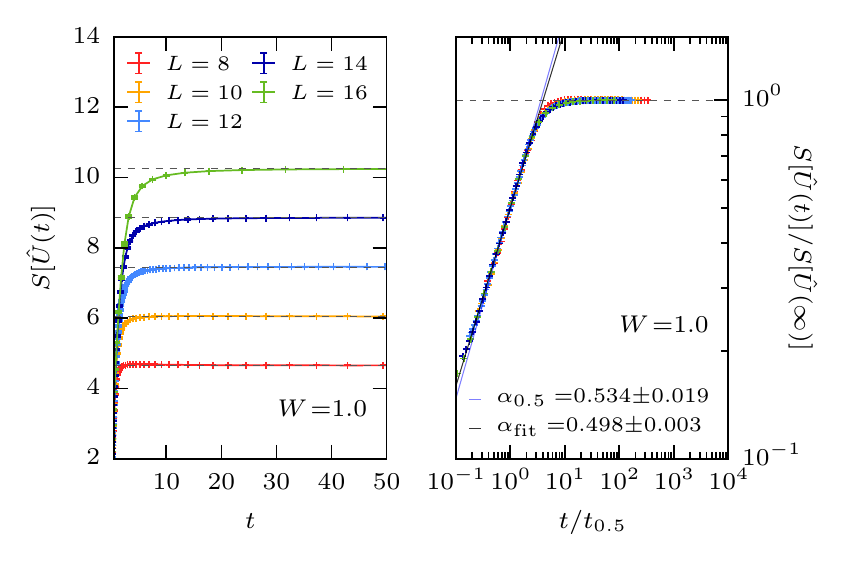}\\[-0.77cm]
    \includegraphics{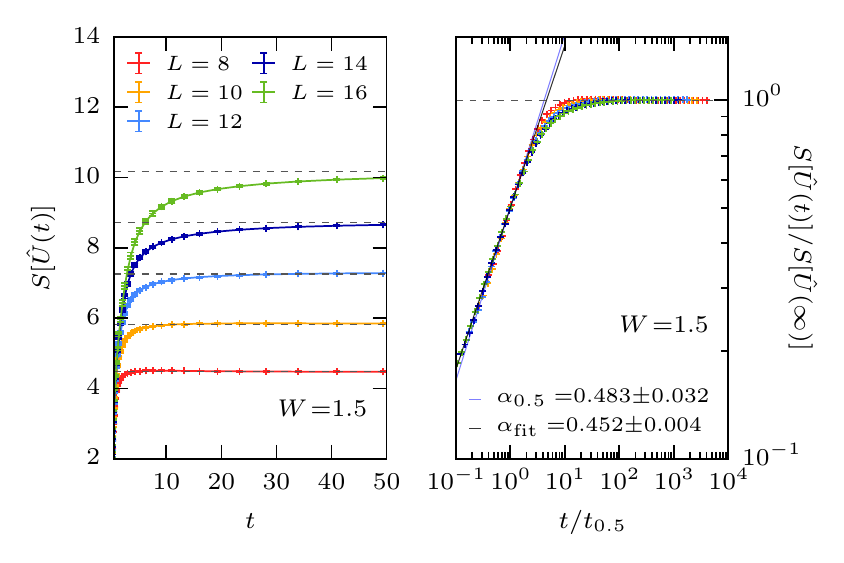}\\[-0.77cm]
    \includegraphics{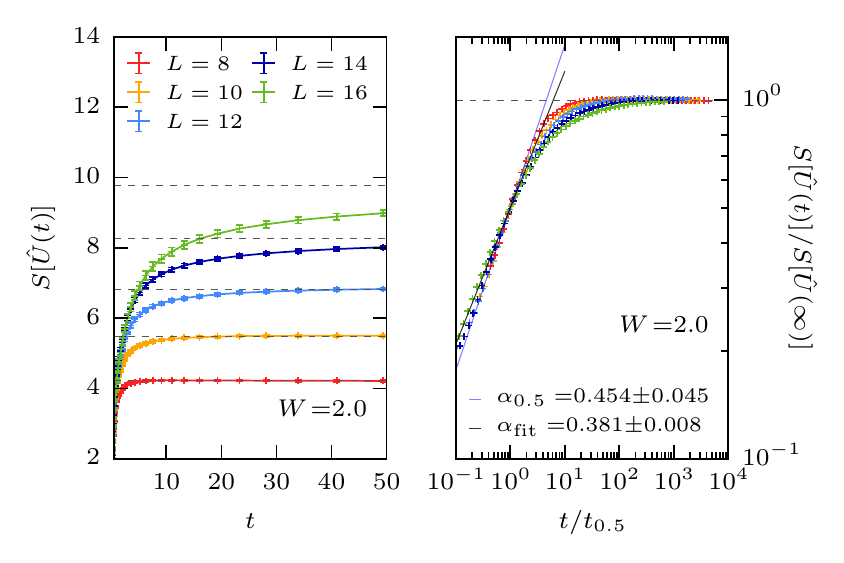}
    \caption{(Left) Growth of the disorder averaged operator entanglement entropy of the evolution operator in the random
        field Heisenberg model ($S_z=0$) at short times for various disorder strengths. (Right)
        Growth of the opEE in units of the saturation value $S_\infty$ (dashed horizontal lines) as a function of time in
        units of the half saturation time $t_{0.5}$. We observe a power law growth at weak disorder
        with an exponent $\alpha<1$, which is valid for intermediate times, when $t\approx t_{0.5}$.
    At stronger disorder, the exponent decreases and finite size effects are stronger, seemingly
leading to a long domain of sup-power law growth of the opEE. }
    \label{fig:opEEscaling}
\end{figure}

In Fig. \ref{fig:opEEscaling}, we show the growth of the disorder averaged opEE for different system sizes and disorder
strengths. The saturation values obtained for very long times are displayed by dashed lines. It is
obvious that the opEE of the evolution operator reaches the saturation value at much later times at
stronger disorder ($W=2$), compared to weak disorder ($W=1$). This is consistent with the numerical
evidence for slow dynamics in this region of the phase diagram, leading to a
subballistic growth of the (state) entanglement entropy after a quench\cite{luitz_extended_2016},
power law information transport as quantified by the out of time order correlation
function\cite{luitz_information_2017} and subdiffusive
transport\cite{bar_lev_dynamics_2014,bar_lev_absence_2015,agarwal_anomalous_2015,luitz_extended_2016,znidaric_diffusive_2016,luitz_anomalous_2016}
(see Ref. \onlinecite{luitz_ergodic_2016} for a recent review of the numerical evidence). 

To analyze the finite size scaling of the opEE of the evolution operator, we conjecture that the
opEE will grow as a power law in time up to saturation, as was observed for the EE after a
quench\cite{luitz_extended_2016}. Then, we can make the scaling assumption that for
hydrodynamic times (after an initial transient, but before the saturation), the opEE grows like
$S[\hat U(t)] = \beta t^\alpha$ and thus we can estimate the time $t_\infty$ after
which the opEE saturates:

\begin{equation}
    t_\infty \propto S( \infty ) ^{\frac{1}{\alpha}},
    \label{eq:tinf}
\end{equation}
where $S_\infty $ is the saturation value $ S[\hat U(\infty)]$

With this natural timescale in the problem, we propose the scaling hypothesis

\begin{equation}
    S[\hat U(t)]/S(\infty) = f(t/t_\infty).
    \label{eq:scalinghypo}
\end{equation}

Numerically, it is difficult to determine the saturation time accurately, as in its proximity the
power law growth seems to be violated. Therefore, we define instead the time $t_{0.5}$ at which the opEE
reaches half of the saturation value by
\begin{equation}
    S[\hat U(t_{0.5})] =\frac{S[\hat U(\infty)]}{2}
    \label{eq:halfsat}
\end{equation}
and use $t_{0.5}$ as the natural timescale. We determine this time by interpolating the time
evolution of the opEE and solving Eq.~\eqref{eq:halfsat} for $t_{0.5}$ numerically for each system
size and disorder strength. The associated errorbar of $t_{0.5}$ is estimated by the error of the
opEE divided by the derivative of the opEE with respect to time. 

In the right panels of Fig.~\ref{fig:opEEscaling}, we display the opEE divided by the saturation
value $S[\hat U(\infty)]$ as a function of time in units of the half saturation time $t_{0.5}$ for
different system sizes on a log-log scale. At weak disorder, all curves collapse almost perfectly to one universal
curve, displaying a clear power law for $t \sim t_{0.5}$. For stronger disorder, finite size effects become more visible
but it seems that the curves still converge to a universal curve for larger system sizes. At a
disorder strength of $W=2$ (bottom panel), the power law regime is shorter than at weaker disorder
and at intermediate times an extended regime of slow growth of the opEE is visible in the curvature
of the curve. Currently, it is unclear where this regime comes from, but we may speculate that it is
due to the fact that at this disorder strength for finite systems a fraction of the eigenstates of
the Hamiltonian are already many-body localized\cite{luitz_many-body_2015}. Although there is no
direct connection to the behavior of the eigenstates of the Hamiltonian,  to the growth of the opEE
of the evolution operator can be influenced by this fact and therefore exhibit slower
dynamics. At weaker disorder, the fraction of localized states in the spectrum of the Hamiltonian is
much smaller and therefore the effect of slower dynamics can be expected to be less visible, which
is the case in our results.

Let us finally address the power law growth of the opEE at short times. In the previous subsection,
we have shown that for a Floquet system, the growth is almost linear, however in the random
Heisenberg chain slower dynamics can be expected. We use two methods to determine the exponent of
the power law growth: First, we fit a power law to the growth of the opEE in time for the largest
system size over a time window where it appears to be linear on a log-log scale, yielding an
exponent $\alpha_\text{fit}$. The fit and the value of the exponent is reported in Fig.~\ref{fig:opEEscaling}. The second approach relies on the scaling ansatz, since according to the
scaling arguments explained above, we expect that 

\begin{equation}
    t_{0.5}  \propto L^{\frac 1 \alpha},
    \label{eq:t05scaling}
\end{equation}
and we can use this ansatz to fit a power law to $t_{0.5}(L)$, yielding $\alpha_{0.5}$. Both
exponents agree reasonably well with errorbars. We also show the corresponding power law curves
together with the data collapse for comparison.




\subsection{Growth of $S[\hat U(t)]$ in the MBL phase}

\begin{figure}[h]
    \centering
    \includegraphics{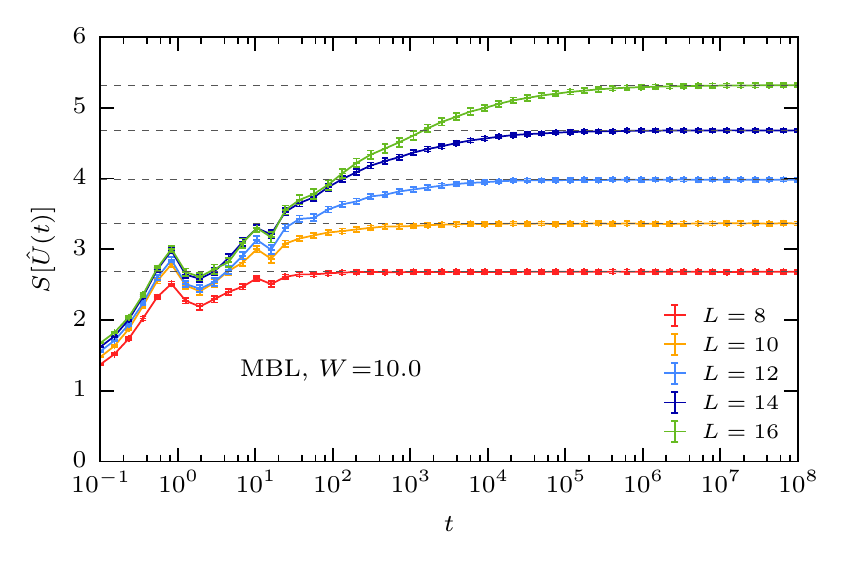}
    \caption{Growth of the operator entanglement entropy in the MBL phase of the random field Heisenberg model ($S_z=0$). }
    \label{fig:mblgrowth}
\end{figure}

It is known that the entanglement dynamics in MBL systems is much slower than in ergodic systems, in
fact after a quench, the (state) entanglement entropy grows logarithmically and saturates to a
volume law value with a suppressed
prefactor\cite{znidaric_many-body_2008,bardarson_unbounded_2012,luitz_extended_2016,singh_signatures_2016}.

In Fig.~\ref{fig:mblgrowth}, we show the disorder averaged opEE of the evolution operator for
different system sizes of the random field Heisenberg chain \eqref{eq:Heisenberg} at strong
disorder $W=10$, where the system is surely in the MBL phase.
On a logarithmic scale in time, it is visible that the saturation value is approached extremely
slowly and our results are consistent with a logarithmic growth of the opEE, although larger system
sizes would be required to test this hypothesis thoroughly. 


\section{Conclusion}
\label{sec:conclusion}
We have defined the opEE of the time evolution operator and analyzed the saturation and growth
patterns in various spin systems. 

The Floquet system is the most chaotic among the models we studied. It has a linear initial growth
of the opEE, saturating at the Page value: the average EE of a random unitary operator. We note that
the linear growth is also observed in other Floquet models\cite{mishra_protocol_2015} and quenches
under a random unitary gate\cite{nahum_quantum_2016}. 
It would be interesting to use the hydrodyanmic theory and surface growing model in Ref. \onlinecite{nahum_quantum_2016} to explain the linear growth in our opEE. 

We also consider another chaotic system with global conservation laws: the Heisenberg model with
disorder field. There, we find a power law growth with an almost perfect data collapse in the weak
disorder regime 
and a saturation value only less than the Page value by an non-extensive amount. We believe that the
conservation law and locality of the interaction is responsible for the slower growth and smaller
saturation value (compared to Floquet model). The opEE in the MBL phase exhibits a logarithmic
growth in time and saturates to an extensive value which is given by a fraction of the saturation
value in chaotic models. 

Due to the mapping to a quench problem in Sec.~\ref{sec:map_quench}, we understand the behavior of opEE through the knowledge of the wave
function EE after a global quench. Yet one advantage of the opEE is its initial state independence.
It is therefore useful to characterize the scrambling properties of the time evolution operator
itself. 

The the channel-state duality in App.~\ref{app:channel_state} gives us a double-system picture that also emerges in the thermofield
double state in the study of the holography with the presence of the eternal black
hole\cite{maldacena_eternal_2003}. The opEE is the state EE of the global quenched dual state, and in the gravity dual, the Ryu-Takayanagi surface\cite{ryu_aspects_2006} will probe behind the horizon of the black hole\cite{hartman_time_2013}. It would be interesting to reproduce the scaling and saturation in a holographic calculation.




{\it Note:} Shortly after the submission of this manuscript, we became aware of a
preprint\cite{dubail_entanglement_2016} that investigates the opEE mainly from the CFT perspective.
Section 4 of Ref. \onlinecite{dubail_entanglement_2016} has an overlap with some our conclusions. 

\begin{acknowledgments}
    TZ benefited from discussions with Mark K. Mezei and Thomas Faulkner about the holographic interpretation of the quench EE. 
    TZ is supported by the National Science Foundation under grant number NSF-DMR-1306011.
    DJL is grateful for discussions with Yevgeny Bar Lev.
    We appreciate Toma{\v z} Prosen's effort of reading the manuscript and providing comments on the
    distinction of COE and CUE and thank Fabien Alet and Yevgeny Bar Lev for their comments on the manuscript. 
    This work was supported in part by the Gordon and Betty Moore Foundation's EPiQS Initiative
    through Grant No. GBMF4305 at the University of Illinois.
    This work made use of the Illinois Campus Cluster, a computing resource that is operated by the
    Illinois Campus Cluster Program (ICCP) in conjunction with the National Center for
    Supercomputing Applications (NCSA) and which is supported by funds from the University of
    Illinois at Urbana-Champaign.
\end{acknowledgments}

\onecolumngrid
\appendix

\section{Channel-State Duality}
\label{app:channel_state}

Here we view the opEE in the light of the channel-state duality originated from the quantum information community. We restrict to the unitary channel that is relevant to the opEE. A more detailed account and application can be found for example in Ref. \onlinecite{de_pillis_linear_1967,choi_completely_1975,jamiolkowski_linear_1972, jiang_channel-state_2013, hosur_chaos_2016}.

For any linear operator expanded in a basis $|i\rangle \in \mathcal{H}$, 
\begin{equation}
\begin{aligned}
U &= \sum_{ij} U_{ij} |i \rangle \langle j  | 
\end{aligned}
\end{equation}
we can always construct a corresponding state $| \psi\rangle$ in the enlarged Hilbert space $\mathcal{H} \times \mathcal{H}$
\begin{equation}
  |\psi \rangle  =   \sum_{ij} |i \rangle \otimes U_{ij} | j \rangle ^* = \sum_{i} |i\rangle  \otimes U |i  \rangle^*.
\end{equation}
Operationally, we just replace the bra $\langle j | $ by a ket $|j\rangle^*$  which is the complex conjugation of the state $|j\rangle$. This choice makes the state $|\psi\rangle$ basis independent, which can be easily verified by applying a unitary transformation $V$
\begin{equation}
\begin{aligned}
|\psi' \rangle  &= \sum_{i'} | i ' \rangle   \otimes U |i'  \rangle^* = \sum_{i' i j} V_{i' i } V^*_{i'j} | i  \rangle   \otimes U |j  \rangle^* \\
&= \sum_{ij} \delta_{ij}  | i  \rangle   \otimes U |j  \rangle^*  = |\psi \rangle. 
\end{aligned}
\end{equation}
So the unique state $\ket{\psi}$ dual to the unitary operator $U$ contains all its information, and one can study this state instead to gain knowledge of the operator. 

\begin{figure}[h]
\centering
\includegraphics[width=0.3\columnwidth]{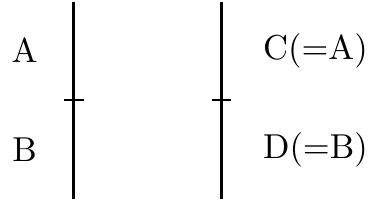}
\caption{Channel state duality point of view of opEE. The vertical lines correspond to the two
copies of the original system and the bipartition of the system into $A$ and $B$ has to be performed
equally in both copies.}
\label{fig:channel.pdf}
\end{figure}

The dual state is defined on two copies of the original system, and the unitary operator is acting only on one of them. Partitioning of the operator corresponds to an identical space partitions in these two copies of system, which is shown in Fig.~\ref{fig:channel.pdf}.
The opEE is then identical to the state EE of the $A$, $B$ partition for the dual state
$\ket{\psi}$. We use this picture to analytically compute the average opEE of a random unitary
operator in App.~\ref{app:aver_rand_op_EE}.




\section{Average opEE of Random Unitary Operator}
\label{app:aver_rand_op_EE}

In this appendix, we prove that the average opEE of random unitary operator (circular unitary
ensemble) is equal to the Page value. \onlinecite{musz_unitary_2013} notices that the distribution of Schmidt eigenvalues of random operator and random state of doubled system are different. However it is argued that in the large system limit, the "reshuffled" matrix should asymptotically follow the same Ginibre ensemble and hence consistent with numerically calculated Page value. We here present a direct mathematical calculation to prove this point. 

We use the standard replica trick and average over the Haar measure $[dU]$ of the unitary group ${\rm U}(N)$ to compute the EE
\begin{equation}
\bar{S}[U] = - \int [dU]\,  \partial_{n}   \tr( \rho^n[U]) \Big|_{n=1}
\end{equation}
and further assume that the derivative and integral commute, so that we can compute the average first 
\begin{equation}
\overline{\tr( \rho^n)} = \int [dU]\,   \tr( \rho^n[U])
\end{equation}

\begin{figure}[h]
\centering
\includegraphics[width=0.6\columnwidth]{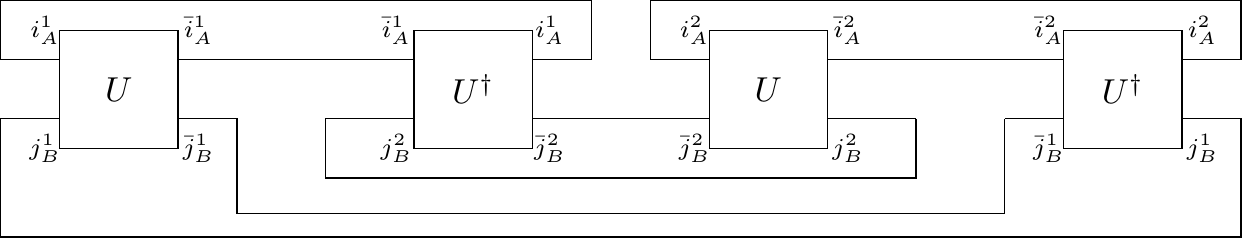}
\caption{Diagrammatic representation of $\tr( \rho^2[U] )$. }
\label{rho_2.pdf}
\end{figure}

In a chosen basis, the matrix element can be written as $U_{i_Aj_B, \bar{i}_A \bar{j}_B}$, where the combination of $i_A$, $j_B$ exhausts the indices for a state (left line of Fig.~\ref{fig:channel.pdf}), and the same for $\bar{i}_A$, $\bar{j}_B$ (right line of Fig.~\ref{fig:channel.pdf}). We need a partial transpose to obtain the expansion coefficient in the operator basis
\begin{equation}
U_{i_Aj_B, \bar{i}_A \bar{j}_B} \rightarrow U_{i_A \bar{i}_A , j_B \bar{j}_B}
\end{equation}
where now $i_A$ and $\bar{i}_A$ are indexing the $A_i$ basis etc. The density matrix for the operator is then
\begin{equation}
\rho[U]_{i_A \bar{i}_A, i'_A \bar{i}'_A} = U_{i_A \bar{i}_A , j_B \bar{j}_B} U^*_{ j_B \bar{j}_B,
i'_A \bar{i}'_A },
\end{equation}
summing over repeated indices.
The diagrammatic representation in Fig.~\ref{rho_2.pdf} can guide\cite{hosur_chaos_2016} us to write down the complicated index structure of $\tr( \rho^n[U])$. 
For a $2^n \times 2^n$ matrix $U_{i_Aj_B, \bar{i}_A \bar{j}_B}$, the upper two closed lines on each block represents $A$ region indices $i_A, \bar{i}_A$ and the lower two closed lines represents $B$ region indices $j_B, \bar{j}_B$. The two ends of connecting lines are contracting indices. So for example, the diagram in Fig.~\ref{rho_2.pdf} can be translated to 
\begin{equation}
\begin{aligned}
\tr( \rho^2[U] ) = &  U_{i_A^1 j^1_B, \bar{i}^1_A \bar{j}^1_B} \left(U^{\dagger}\right)_{\bar{i}^1_A  \bar{j}^2_B, i^1_A  j_B^2 }  U_{i^2_A j_B^2, \bar{i}^2_A \bar{j}^2_B} \left(U^{\dagger}\right)_{\bar{i}^2_A  \bar{j}^1_B, i^2_A  j^1_B } \\
=&  U_{i_A^1 j^1_B, \bar{i}^1_A \bar{j}^1_B} U^*_{i^1_A  j_B^2, \bar{i}^1_A  \bar{j}^2_B } U_{i^2_A j_B^2, \bar{i}^2_A \bar{j}^2_B} U^*_{ i^2_A  j^1_B , \bar{i}^2_A  \bar{j}^1_B}
\end{aligned}
\end{equation}
where $*$ represents the complex conjugate of the indexed element. 

\begin{figure}[h]
\centering
\includegraphics[width=0.8\columnwidth]{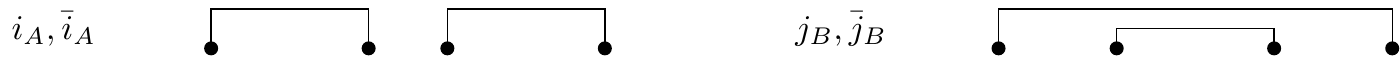}
\caption{Eight indices in $\tr(\rho^2[U])$, where contractions are performed for the 4 pairs. }
\label{lid.pdf}
\end{figure}

The same type of integral also appears in the discussion of Haar scrambling in
Ref.~\onlinecite{shenker_black_2014} and \onlinecite{hosur_chaos_2016}, where the $n = 2$ case is
calculated by the Weingarten formula to obtain the R\'enyi entropy. We here apply the general Weingarten formula for the integration on the unitary group, 
\begin{equation}
\label{eq:tr_rho_2}
\begin{aligned}
\int [dU] &U_{i_1,j_1} U_{i_2,j_2}  \dots  U_{i_n, j_n}  U^*_{i'_1,j'_1} U^*_{i'_2,j'_2}  \dots  U^*_{i'_n, j'_n}  \\
& = \sum_{ \sigma, \tau \in S_n} \delta_{i_1 i'_{\sigma(1)}}\delta_{i_2 i'_{\sigma(2)}} \dots \delta_{i_n i'_{\sigma(n)}} \delta_{j_1 j'_{\tau(1)}}\delta_{j_2 j'_{\tau(2)}} \dots \delta_{j_n j'_{\tau(n)}} {\rm Wg}(N, \sigma \tau^{-1} )  \\
\end{aligned}
\end{equation}
where the sum is taken over all possible permutations in $S_n$, and $N$ is the size of the matrix
$2^L$. $\rm Wg$ is the Weingarten function (see detailed definition and the first few examples in
Ref. \onlinecite{collins_integration_2006}), whose large $N$ limit (thermodynamic limit)\cite{collins_integration_2006} is given by
\begin{equation}
\begin{aligned}
  {\rm Wg}(N, \sigma ) &= \frac{1}{N^{n + |\sigma|} } \Big[ \prod_{\sigma = C_1 C_2 \cdots C_k } (-1)^{|C_i|-1} {\rm Catalan}_{|C_i|}  + \mathcal{O}(N^{-2}) \Big]
\end{aligned}
\end{equation}
where the $C_i$ are the cycle decomposition of $\sigma$ , $|C_i|$ are the number of elements in this cycle, Catalan is the Catalan number, $|\sigma|$ is its Cayley distance to the identity (minimal number of transpositions that makes it identity). Obviously, the dominant term in $N \rightarrow \infty$ limit is the one with $\sigma = e$
\begin{equation}
{\rm Wg} ( N, e ) = \frac{1}{N^n} + \mathcal{O}(N^{-n-2})
\end{equation}
Consequently, in the integration of the $U$ only $\sigma = \tau$ are relevant, i.e. terms whose $i$ index and $j$ index share the same permutation
\begin{equation}
\int [d U] \cdots \sim \frac{1}{N^{n}}\sum_{ \sigma \in S_n} \delta_{i_1 i'_{\sigma(1)}}\delta_{i_2 i'_{\sigma(2)}} \dots \delta_{i_n i'_{\sigma(n)}} 
\delta_{j_1 j'_{\sigma(1)}}\delta_{j_2 j'_{\sigma(2)}} \dots \delta_{j_n j'_{\sigma(n)}} 
\end{equation}
The contractions of these delta function can be converted to loop counting in planar diagrams. Let us illustrate the example of $n = 2$ 
\begin{equation}
\label{eq:tr_rho_2}
\int [dU] \tr( \rho^2[U] )  =  \int [dU]\,  U_{i_A^1 j^1_B, \bar{i}^1_A \bar{j}^1_B} U^*_{i^1_A  j_B^2, \bar{i}^1_A  \bar{j}^2_B }
U_{i^2_A j_B^2, \bar{i}^2_A \bar{j}^2_B} U^*_{ i^2_A  j^1_B , \bar{i}^2_A  \bar{j}^1_B}
\end{equation}
where the four indices may be represented as the lids in Fig.~\ref{lid.pdf}
After doing the integration, the delta functions for each permutation element $\sigma$ will close these diagrams. For example when $\sigma = (12)$, there are $1\times 2$ loops for A indices and $2\times 2$ loops for B indices, and then the corresponding factor is $2^{2\ell_A + 4\ell_B}$. 
\begin{figure}[h]
\centering
\includegraphics[width=0.8\columnwidth]{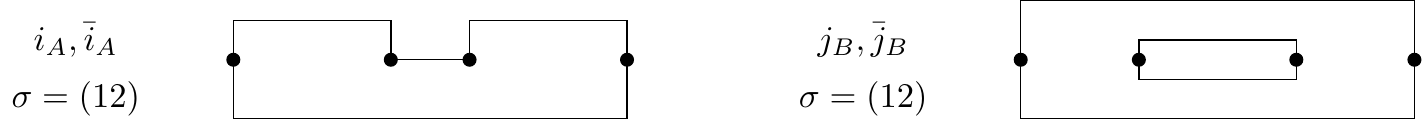}
\caption{Delta functions for each permutation element $\sigma$ will close these diagrams. Each loop will contribute a $(2^{\ell_A})^2$ or $(2^{\ell_B})$, where the square is for two copies of indices.}
\label{loop.pdf}
\end{figure}

The loop counting is a combinatorial problem that can be formulated in terms of its generating function (by trial and error). Let
\begin{equation}
f_n(x,y)= \sum_{g \in S_n} x^{\chi( g )} y^{\chi ( \tau g ) }
\end{equation}
where $\chi( g)$ is the number of cycles in the permutation and $\tau = (12\cdots n)$. The average of the trace can be expressed by this polynomial 
\begin{equation}
\overline{\tr( \rho^n[U] )} =  \frac{1}{(2^{2\ell_A + 2\ell_B} )^n}f_n( 4^{\ell_A}, 4^{\ell_B})  = \frac{1}{4^{nL}}f_n( 4^{\ell_A}, 4^{\ell_B})
\end{equation}

At this point, we can apply Page's state result as a shortcut. For a random state, the component of the wavefunction is
\begin{equation}
\psi_{\bar{i}_A \bar{j}_B } =  U_{1, \bar{i}_A \bar{j}_B } 
\end{equation}
where $U$ is again taken from the Haar measure. To contrast, we write down the state version of integral for $n = 2$
\begin{equation}
 \int [dU] \tr( \rho^2[\psi] )  =  \int [dU]\, U_{1, \bar{i}^1_A \bar{j}^1_B} U^*_{1, \bar{i}^1_A  \bar{j}^2_B }
U_{1, \bar{i}^2_A \bar{j}^2_B} U^*_{ 1 , \bar{i}^1_A  \bar{j}^1_B}
\end{equation}
The whole process using the Weingarten formula and its asymptotics can be similarly applied; the only difference is that the state has no unbarred set of indices
\begin{equation}
\overline{\tr(\rho^n[\psi] )} = \frac{1}{2^{nL}}f_n( 2^{\ell_A}, 2^{\ell_B} ) 
\end{equation}
so the opEE will be the Page value of a state with length $2L$ and partition $2 \ell_A + 2 \ell_B= 2L$, \ie
\begin{equation}
S[U] = 2 \ell_A \ln 2 - 2^{2\ell_A - 2\ell_B - 1} 
\end{equation}

We also manage to do a direct combinatorial computation for equal partition, where the top
coefficient of the generating function\cite{stanley_catalan_2015} $a_n$ (cf. Ref. \onlinecite{dulucq_combinatorial_1998} for why the top power is $n+1$ and the concept of genus)
\begin{equation}
f_n( x, x ) = \sum_{g \in S_n}  x^{\chi( g )  + \chi ( \tau g ) } = a_n x^{n+1} + \cdots
\end{equation}
determines the trace
\begin{equation}
\overline{\tr(\rho^n[\psi] )} = \frac{1}{4^{nL}}f_n( 4^{\frac{L}{2}}, 4^{\frac{L}{2}} ) = \frac{a_n }{2^{(n-1) L}} + \mathcal{O}(\frac{1}{2^{nL} }). 
\end{equation}
By analytic continuation,  the EE is
\begin{equation}
S[U] = L \ln 2 -  \partial_n a_n \Big|_{n=1} + \mathcal{O}(\frac{1}{2^{L} }).
\end{equation}
Through a series of bijections, one can show that $a_n$ is the Catalan number(exercise 118
of Refs. \onlinecite{stanley_catalan_2015}, \onlinecite{simion_noncrossing_2000})
\begin{equation}
a_n =  \frac{2n!}{n!(n+1)!}
\end{equation}
Therefore
\begin{equation}
\partial_n \frac{2n!}{n!(n+1)!}\Big|_{n = 1}  = \partial_n \frac{1}{n(n+1) B( n, n+ 1)}\Big|_{n = 1} = \frac{1}{2}
\end{equation}
gives us the correct deficit of the Page value
\begin{equation}
S = L \ln 2 - \frac{1}{2}
\end{equation}




\section{Lin Table Algorithm for $S_z = 0$ Sector}
\label{app:lin_table}

We take the $\sigma_z$ basis in the Hilbert space, each of which are eigenvectors of total $\sum_i
\sigma^z_i $. Consider the subspace where the eigenvalue of the total $S^{z}$ is zero. Each basis
state then has an equal number of $\uparrow$ spins and $\downarrow$ spins. In a $2N$ sites system, the dimension of the subspace is ${2N \choose N}$. 

For any state in this subspace, we can take advantage of the constraint to do a partial Schmidt decomposition
\begin{equation}
|\psi \rangle = \sum_{n_{\up} } \sum_{ij} \psi_{n_{\up} }^{ij} |n_{\up}, i \rangle_A |  n -  n_{\up}, j \rangle_B 
\end{equation}
where $\psi^{ij}_{n_{\up}}$ is a block diagonal matrix, the number of up spins $n_{\up}$ in part A is the block index and $ij$ are the row and column indices within the block. The dimension of each block is ${ N \choose n_{\up}}^2 $, and the identity
\begin{equation}
\sum_{n_{\up} = 0 }^N { N \choose n_{\up}}^2  = { 2N \choose N } 
\end{equation}
ensure that the coefficients from the $\sigma^z$ basis wavefunction to the block elements $\psi^{ij}_{n_{\up}}$ is just a permutation. 

We use a 2N-bit binary number to represent the $\sigma^z$ basis. In the example of $N = 3$, the $\sigma^z$ basis wavefunction wavefunction elements are
\begin{equation}
\begin{pmatrix}
             & 000 & 001 & 010 & 011 & 100 & 101 & 110 & 111 & \\ \hline
000\vline    & & & & & & & & \psi_1  \\
001\vline    & & & & \psi_2 & & \psi_3  & \psi_4 \\
010\vline    & & & & \psi_5 & & \psi_6  & \psi_7 \\
011\vline    & & \psi_8 & \psi_9 &  & \psi_{10} \\
100\vline    & & & & \psi_{11} & & \psi_{12}  & \psi_{13} \\
101\vline    & & \psi_{14} & \psi_{15} &  & \psi_{16} \\
110\vline    & & \psi_{17} & \psi_{18} &  & \psi_{19} \\
111\vline    & \psi_{20} \\
\end{pmatrix}
\end{equation}
and the corresponding $\psi^{ij}_{n_{\up}}$ matrix is 
\begin{equation}
\begin{pmatrix}
             & 111 & (011 & 101 & 110) & (001 & 010 & 100) & 000 & \\ \hline
000\vline    & \psi_1 \\
001\vline    &     & \psi_2 & \psi_3  & \psi_4 \\
010\vline    &     & \psi_5 & \psi_6  & \psi_7 \\
100\vline    &     & \psi_{11} & \psi_{12}  & \psi_{13} \\
011\vline    & & & & &\psi_{8} & \psi_{9} &  \psi_{10} \\
101\vline    & & & & &\psi_{14} & \psi_{15} & \psi_{16} \\
110\vline    & & & & &\psi_{17} & \psi_{18} & \psi_{19} \\
111\vline    & & & & & & & & \psi_{20} \\
\end{pmatrix}
\end{equation}
If we store $\psi^{ij}_{n_{\up}}$ in a row vector, then the index of $\psi$ will be 
\begin{equation}
(1,2,3,4,5,6,7,11,12,13,8,9,10,14,15,16,17,18,19,20)
\end{equation} 
The permutation element we are looking for in this $n = 3$ example is $(8,11),(9,12),(10,13)$. The
algorithm needs to figure out the conversion table from the $\sigma^z$ basis elements to the block
elements and then do the Schmidt decomposition for each blocks, which is much more efficient than
doing it in the full Hilbert space. We note that this can be done efficiently using the method of
Lin tables as pointed out by H. Lin in Ref. \onlinecite{lin_exact_1990}.





\end{document}